\documentclass[12pt]{article}
\usepackage{epsfig}

\usepackage{slashed}
\newcommand{\notp}{{\slashed{p}}}

\usepackage{amsmath}

\begin{document}

\title{
\vskip-3cm{\baselineskip14pt
\centerline{\normalsize DESY 08--156\hfill ISSN 0418-9833}
\centerline{\normalsize NYU--TH/09/02/20\hfill}
\centerline{\normalsize February 2009\hfill}}
\vskip1.5cm
\bf On-shell renormalization of the mixing matrices in Majorana neutrino
theories}

\author{Andrea A. Almasy$^*$, Bernd A. Kniehl$^*$ and Alberto Sirlin$^\dagger$\\
\\
{\normalsize\it $^*$ II. Institut f\"ur Theoretische Physik, Universit\"at
Hamburg,}\\
{\normalsize\it Luruper Chaussee 149, 22761 Hamburg, Germany}\\
\\
{\normalsize\it $^\dagger$ Department of Physics, New York University,}\\
{\normalsize\it 4 Washington Place, New York, New York 10003, USA}}

\date{}

\maketitle

\begin{abstract}
We generalize a recently proposed on-shell approach to renormalize the Cabibbo-Kobayashi-Maskawa quark-mixing matrix to the case of an extended leptonic sector that includes Dirac and Majorana neutrinos in the framework of the seesaw mechanism. 
An important property of this formulation is the gauge
independence of both the renormalized and bare lepton mixing matrices. 
Also, the texture zero in the neutrino mass matrix is preserved.

\medskip

\end{abstract}

\newpage

\section{Introduction}

Renormalizability endows the Standard Model (SM) with enhanced predictive
power due to the fact that ultraviolet (UV) divergences from quantum effects
can be eliminated by a redefinition of a finite number of independent
parameters, such as masses and coupling constants. 
Furthermore, it has been known for a long time that, in the most frequently
employed formulations in which the complete bare mass matrices of quarks are
diagonalized, the Cabibbo-Kobayashi-Maskawa (CKM) quark mixing matrix
\cite{ref:5} must be also renormalized \cite{ref:6}. In fact, this problem has been the object of several interesting studies over the last two decades \cite{ref:7,ref:1}.

A matter of considerable interest is the generalization of these considerations to minimal renormalizable extensions of the SM. In particular, in Refs.~\cite{ref:10,ref:11} the mixing-matrix
renormalization was extended to theories that include isosinglet neutrinos and admit the presence of lepton-number-violating Majorana masses. 
A minimal realization of such a theory is the SM with right-handed Dirac and Majorana neutrinos \cite{ref:10,ref:12}, an appealing scenario that may explain the smallness of the observed neutrino masses and may lead to neutrino-less double beta decays. Furthermore, this minimal extension may give rise to a number of observable phenomena, such as lepton-flavor and/or lepton-number violation in $\mu$, $\tau$ \cite{ref:13} and $Z$-boson decays \cite{ref:16}, or
to possible lepton-number-violating signals at high-energy colliders
\cite{ref:17}.

The aim of this paper is to generalize the on-shell renormalization of the CKM matrix
recently proposed in Ref.~\cite{ref:1} to extensions of the SM in which
the lepton sector contains Majorana neutrinos. 
An important property is that this formulation complies with UV finiteness and gauge independence,\footnote{%
Throughout this paper, the term {\it gauge independence} is used as an
abbreviation for {\it gauge parameter independence}.}
 and also preserves the basic structure of the theory. 
In particular, the texture zero ($m_L^{\prime0}=0$) in the neutrino mass
matrix is preserved by renormalization.

This paper is organized as follows.
After briefly reviewing in Section~\ref{sec:majorana} the basic formalism of the
seesaw mechanism in the minimal extension of the SM neutral-lepton sector, we
evaluate in Section~\ref{sec:self-energies} the one-loop self-energy insertions
(see Figs.~\ref{charged-lepton:diag} and \ref{majorana:diag}) in an external
charged-lepton or Majorana-neutrino leg, perform the separation into
wave-function renormalization (wfr) and self-mass (sm) amplitudes, and show
explicitly the cancellation of gauge dependences in the latter. 
As in the quark case \cite{ref:1}, the mass counterterm matrix, to be
discussed in Section~\ref{sec:mass-count}, is chosen to cancel, as much as
possible, the sm contributions. 
In Section~\ref{sec:mixing-count}, we discuss the diagonalization of the complete
mass matrix and show explicitly how this procedure generates mixing
counterterm matrices in a manner that preserves the basic structure of the
theory, as well as gauge independence and UV finiteness. 
Finally, our conclusions are summarized in Section~\ref{sec:conclusions}.  

\section{Neutrino see-saw mechanism}
\label{sec:majorana}

We consider a minimal, renormalizable extension of the SM, based on the
$SU(2)_I\otimes U(1)_Y$ gauge group, that can naturally accommodate heavy
Majorana neutrinos.
We allow for an arbitrary number $N_G$ of fermion generations.
Similarly to the SM, each lepton family contains one weak-isospin ($I$)
doublet $(\nu_{L,i}^{\prime 0},l_{L,i}^{\prime 0})$ of left-handed states with
weak hypercharge $Y=-1$ and one right-handed charged-lepton state
$l_{R,i}^{\prime 0}$ with $I=0$ and $Y=-2$ ($i=1,2,\ldots,N_G$). 
In addition, there is a total of $N_R$ right-handed neutrinos
$\nu_{R,i}^{\prime 0}$ with $I=Y=0$ ($i=1,2,\ldots,N_R$). 
The superscript 0 denotes bare quantities, while the primes are to remind us that we are dealing with weak-interaction
eigenstates.

The bare Lagrangian density contains the neutrino mass terms
\begin{equation}
{\cal L}^{\prime 0,\nu}=-\frac{1}{2}\left(\overline\nu^{\prime 0}_L,
\overline\nu^{\prime 0C}_R\right)m^{\prime 0,\nu}
\left(\begin{array}{l} \nu^{\prime 0C}_L\\ \nu^{\prime 0}_R\end{array}\right)
+{\rm h.c.},
\label{lagrangian:nu}
\end{equation}
where $\nu^{\prime 0}_L=\left(\nu^{\prime 0}_{L,1},\ldots,
\nu^{\prime 0}_{L,N_G}\right)^T$,
$\nu^{\prime 0}_R=\left(\nu^{\prime 0}_{R,1},\dots,
\nu^{\prime 0}_{R,N_R}\right)^T$, the superscript $C$ denotes charge
conjugation, $T$ means transpose, and $m^{\prime 0,\nu}$ is a complex,
symmetric matrix of the form
\begin{equation}
 m^{\prime 0,\nu}=\left(\begin{array}{ll} m^{\prime 0}_L & m^{\prime 0}_D\\
m^{\prime 0T}_D & m^{\prime 0}_M\end{array}\right).
\label{neutrinomass}
\end{equation}

Unless the SM Higgs sector is supplemented by additional weak-isospin singlets and/or triplets of Higgs fields, invariance under $SU(2)_I\times U(1)_Y$ leads to
$m^{\prime 0}_L=0$. 
In the following, we do assume that $m^{\prime 0}_L=0$.
This allows for the implementation of the seesaw mechanism.

The neutrino mass matrix~(\ref{neutrinomass}) can always be diagonalized
through a unitary transformation.
For the reader's convenience, we present a simple proof in
Appendix~\ref{sec:appendix2}.
The non-negative diagonal matrix then contains the bare neutrino mass
eigenvalues.
The corresponding mass eigenstates are given by
\begin{equation}
\left(\begin{array}{l} \nu^{\prime 0}_L \\
\nu^{\prime 0C}_R\end{array}\right)_a=\sum_b U_{ab}^{0,\nu *}\nu_{L,b}^0,\qquad
\left(\begin{array}{l} \nu^{\prime 0C}_L \\
\nu^{\prime 0}_R\end{array}\right)_a=\sum_b U_{ab}^{0,\nu}\nu_{R,b}^0,
\label{masseigenstates}
\end{equation}
with $a,b,c=1,2,\ldots,N_G+N_R$. It is important to note that Eq.~(\ref{masseigenstates}) leads to the relation
\begin{equation}
\nu_R^0=\nu_L^{0C}.
\label{majoRLC}
\end{equation}
This implies that the bare neutrino mass eigenstates $\nu_L^0$ and $\nu_R^0$ can be identified with the left and right-handed components of the Majorana fields.
\begin{equation}
\nu^0=\nu_L^0+\nu_L^{0C}=\nu_R^{0C}+\nu_R^0.
\label{majoRLC1}
\end{equation}
In Eqs.~(\ref{masseigenstates})--(\ref{majoRLC1}) the first $N_G$ mass eigenstates are identified with the ordinary light
neutrinos (assuming that $N_G=3$), and the remaining $N_R$ states represent
the new neutral leptons predicted by the theory. For convenience, in what follows we denote the charged-lepton mass eigenstates using indices $i,j,k,\ldots$ and the Majorana-neutrino mass eigenstates using indices from the beginning of the alphabet $a,b,c,\ldots$. Accordingly, sums over repeated charged-lepton indices $i,j,k,\ldots$ run from $1$ to $N_G$, while those over the neutrino indices $a,b,c,\ldots$ extend from $1$ to $N_G+N_R$. 
 
The parts of the bare Lagrangian describing the couplings of the $W^\pm$, $Z$, and
Higgs ($H$) bosons to the charged-lepton mass eigenstates, $l_i^0$, and Majorana-neutrino mass eigenstates, $\nu_a^0$, are given by:\footnote{In Eq.~(\ref{Wlnu:int}) we have not included the terms describing the interactions of the neutral bosons with the charged leptons, since they are the same as in the SM.}
\begin{eqnarray}
{\cal L}_W^0 &=& -\frac{g^0}{\sqrt{2}}(W_\mu^-)^0\sum_{i,a}\overline{l}^0_iB^0_{ia}
\gamma^\mu a_-\nu_a^0 +{\rm h.c.},\nonumber\\
{\cal L}_{\phi^\pm}^0&=&-\frac{g^0}{\sqrt{2}m_W^0}(\phi^-)^0\sum_{i,a}\overline{l}_i^0B^0_{ia}
(m_i^0a_--m_a^0a_+)\nu_a^0 +{\rm h.c.},\nonumber\\
{\cal L}_Z^0&=&-\frac{g^0}{4c_w^0}Z_\mu^0\sum_{a,b}\overline{\nu}_a^0\gamma^\mu
(C_{ab}^0a_--C_{ab}^{0*}a_+)\nu_b^0,\nonumber\\
{\cal L}_{\phi^0}^0&=&\frac{ig^0}{4m_W^0}(\phi^0)^0\sum_{a,b}\overline{\nu}_a^0
\left[(m_a^0C_{ab}^{0*}+m_b^0C_{ab}^0)a_+-(m_a^0C_{ab}^0+m_b^0C_{ab}^{0*})a_-)\right]\nu_b^0,
\nonumber\\
{\cal L}_H^0&=&-\frac{g^0}{4m_W^0}H^0\sum_{a,b}\overline{\nu}_a^0
\left[(m_a^0C_{ab}^{0*}+m_b^0C_{ab}^0)a_++(m_a^0C_{ab}^0+m_b^0C_{ab}^{0*})a_-)\right]\nu_b^0,
\label{Wlnu:int}
\end{eqnarray}
where $g$ is the $SU(2)_L$ gauge coupling, $c_w$ the cosine of the electroweak mixing angle, $\phi^\pm$ and $\phi^0$ are the charged and neutral Higgs-Kibble ghosts,
respectively, and $a_\pm=(1\pm\gamma_5)/2$ are the chiral projectors. 
$B$ and $C$ are $N_G\times(N_G+N_R)$ and $(N_G+N_R)\times(N_G+N_R)$
non-unitary matrices, respectively. The bare matrices are defined by the expressions
\begin{equation}
 B_{ia}^0=\sum_kV_{ik}^{0,l}U_{ka}^{0,\nu *}, \qquad
C_{ab}^0=\sum_cU_{ac}^{0,\nu T}U_{cb}^{0,\nu *},
\end{equation}
where $V^{0,l}$ is the unitary $N_G\times N_G$ matrix relating the
weak-interaction and mass eigenstates of the charged leptons and $U^{0,\nu}$ is
the unitary $(N_G+N_R)\times (N_G+N_R)$ matrix relating the corresponding
neutrino eigenstates, defined in Eq.~(\ref{masseigenstates}). 
They obey a number of basic identities, which ensure the
renormalizability of the theory, namely \cite{ref:10,ref:12}
\begin{align} 
\sum_c& B_{ic}^0B_{jc}^{0*} =\delta_{ij},& \sum_i& B_{ia}^{0*}B_{ib}^0=C_{ab}^0,
\label{cond:B}\\
\sum_c& B_{ic}^0C_{ca}^0=B_{ia}^0,& \sum_c& C_{ac}^0C_{cb}^0=C_{ab}^0=C_{ab}^{0\dagger},
\label{cond:C}\\
\sum_c& m_c^0B_{ic}^0B_{jc}^0=0,& \sum_c& m_c^0B_{ic}^0C_{ac}^0=0,&
\sum_c& m_c^0C_{ac}^0C_{bc}^0=0.
\label{cond:mass}
\end{align}
The last three relations are manifestations of the presence of lepton-number
violation in the neutrino sector.

\section{Self-energy corrections to an external leg}
\label{sec:self-energies}

Following the approach of Ref.~\cite{ref:1}, the analysis of external-leg
corrections leads to two classes of contributions:
\begin{itemize}
\item[(i)] terms proportional to the virtual-fermion propagator
$i/(\notp-m_{f^\prime})$ with gauge-indepen\-dent cofactors not involving
$\notp$, where $m_{f^\prime}$ stands generically for the mass of the virtual
fermion;
\item[(ii)] terms in which the virtual propagator is cancelled in
both the diagonal and off-diagonal amplitudes.
\end{itemize}
The gauge-independent cofactors of class (i) and the contributions of class
(ii) are identified with the sm and wfr contributions, respectively.
In analogy with QED, the latter contain both gauge-dependent and UV-divergent
parts but, in the evaluation of physical amplitudes, these pieces cancel the
corresponding contributions from the proper vertex diagrams.
On the other hand, also in analogy with QED, the UV-divergent sm contributions
are cancelled by the UV-divergent parts of the mass counterterms.

In order to implement the analysis of the external-leg corrections, we
evaluate the contributions of Figs.~\ref{charged-lepton:diag} and
\ref{majorana:diag} in the $R_\xi$ gauges and, applying the algorithm
developed in Ref.~\cite{ref:1}, we separate them into sm and wfr amplitudes.
We do not enter into details, but rather present the results and emphasize the
differences with respect to the quark case. 
We first treat the case of an outgoing on-shell charged lepton in
Section~\ref{sec:charged-lepton-SE} and then that of an outgoing on-shell
Majorana neutrino in Section~\ref{sec:majorana-SE}. 
We have chosen to do so, since the charged-lepton case is very similar to that
of quarks, while in the Majorana-neutrino case additional interactions
involving flavor mixing appear. For completeness, in Section~\ref{incominglepton} we discuss also the case of incoming charged leptons and Majorana neutrinos.  

\subsection{Outgoing charged lepton}
\label{sec:charged-lepton-SE}

\begin{figure}[t]
\begin{center}
\includegraphics{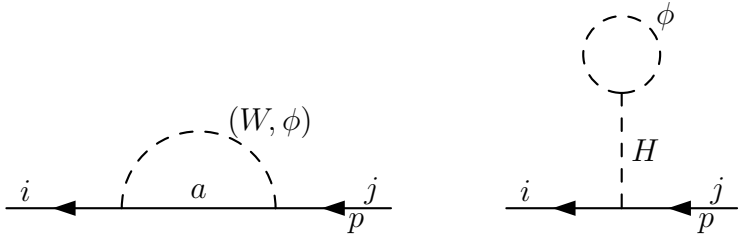}
\caption{Charged-lepton self-energy diagrams at one loop.}
\label{charged-lepton:diag}
\end{center}
\end{figure}

If $i$ is an outgoing on-shell charged lepton, the external-leg amplitude is
obtained by multiplying the diagrams in Fig.~\ref{charged-lepton:diag} on the
left by $\overline{u}_i(p)$, the spinor of the outgoing charged lepton, and on
the right by $i(\notp-m_j)^{-1}$, the propagator of the
virtual charged lepton. 
Thus, the relevant amplitude associated with the external leg is:
\begin{equation}
\Delta {\cal M}_{ij}^{\rm leg}=\overline{u}_i(p) M_{ij}^{(1)}
\frac{i}{\notp-m_j},
\end{equation}
where $M_{ij}^{(1)}$ denotes the contributions of Fig.~\ref{charged-lepton:diag}.

The sm contributions to the external-leg corrections for an outgoing on-shell
charged lepton are:
\begin{eqnarray}
\Delta{\cal M}_{ij}^{{\rm leg,sm}}&=&\frac{g^2}{32\pi^2}\overline{u}_i(p)
\sum_a B_{ia}B_{aj}^\dagger\bigg\{m_i\left(1+\frac{m_i^2}{2m_W^2}\Delta_W
\right)\nonumber \\
&&{}+\left[m_ia_-+m_{j}a_++\frac{m_im_{j}}{2m_W^2}\left(m_ia_++m_{j}a_-\right)
\right]\nonumber\\ 
&&{}\times\left[I(m_i^2,m_W,m_a)-J(m_i^2,m_W,m_a)\right]  \nonumber \\
&&{}-\frac{m_a^2}{2m_W^2}\left(m_ia_-+m_{j}a_+\right)  \nonumber \\ 
&&{}\times\left[3\Delta_W+I(m_i^2,m_W,m_a)+J(m_i^2,m_W,m_a)\right]\bigg\}
\frac{1}{\notp-m_{j}}. 
\label{sm:lepton}
\end{eqnarray}
The functions $I$ and $J$ as well as the UV divergence $\Delta_W$ are defined
in Appendix~\ref{sec:appendix}. 
Note that Eq.~(\ref{sm:lepton}) is a multiple of the virtual charged-lepton
propagator $i(\notp-m_j)^{-1}$ with a cofactor that is gauge
and momentum independent. 
As expected in a chiral theory, it involves the chiral projectors. 
At this point, we should emphasize that Eq.~(\ref{sm:lepton}) is the same as
that for up-type quarks, given in Eq.~(29) of Ref.~\cite{ref:1}, up to
particle changes. 
The only difference is that now no complications due to imaginary parts
appear. 
The amplitudes $I(p^2,m_1,m_2)$ and $J(p^2,m_1,m_2)$ may have absorptive parts
only when their arguments fulfill the condition $p^2>(m_1+m_2)^2$.
In the present case, we have $p^2=m_i^2$, $m_1=m_W$, and $m_2=m_a$, which
ensures that the above inequality can not be satisfied, since the
external-charged-lepton mass is much smaller than that of the $W$ boson.

\begin{figure}[t]
\begin{center}
\includegraphics{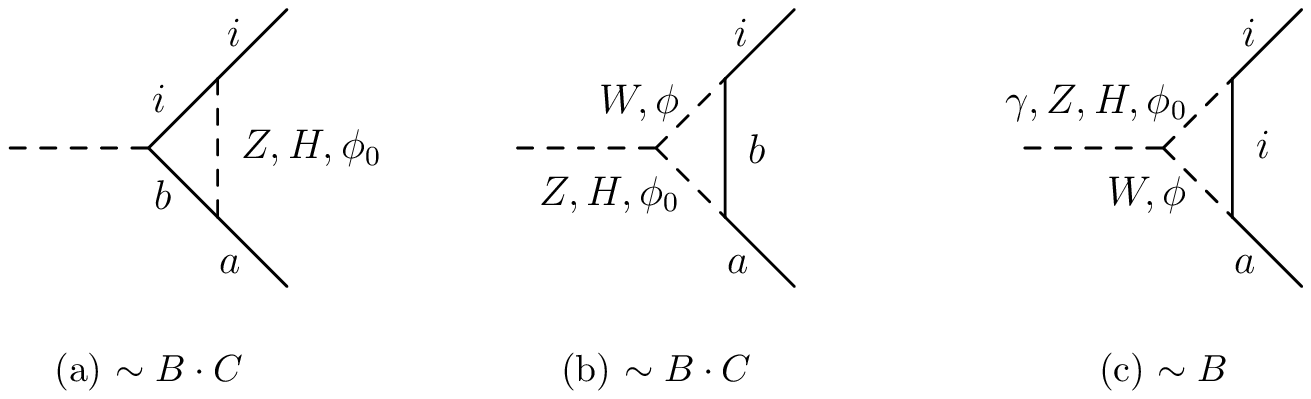}
\caption{Proper $Wl\nu$ vertex diagrams at one loop.}
\label{proper:diag}
\end{center}
\end{figure}

The wfr contributions to the external-leg correction for an outgoing on-shell
charged lepton are given by
\begin{eqnarray}
\Delta{\cal M}_{ij}^{{\rm leg,wfr}}&=&\frac{g^2}{32\pi^2}\overline{u}_i(p)
\sum_a B_{ia}B_{aj}^\dagger\bigg\{\left[I(m_i^2,m_W,m_a)-J(m_i^2,m_W,m_a)
\right]a_+\nonumber\\ 
&&{}+\frac{1}{2m_W^2}\left[m_im_{j}a_-+m_a^2a_+\right]
\left[\Delta_W+I(m_i^2,m_W,m_a)-J(m_i^2,m_W,m_a)\right]\nonumber\\
&&{}-\delta_{ij}\frac{m_i^2m_a^2}{2m_W^2}
\left[I'(m_i^2,m_W,m_a)+J'(m_i^2,m_W,m_a)\right]\nonumber \\
&&{}+\delta_{ij}m_i^2\left(1+\frac{m_i^2}{2m_W^2}\right)
\left[I'(m_i^2,m_W,m_a)-J'(m_i^2,m_W,m_a)\right]\nonumber \\
&&{}+\left[1+\xi_W\left(\Delta_W-\frac{1}{2}+\frac{1}{2}\ln\xi_W\right)\right]
a_+-N(m_W,m_i,m_a,\xi_W)a_+\bigg\}.
\label{wfr:lepton}
\end{eqnarray}
Here $I'(m_i^2,m_W,m_a)$ and $J'(m_i^2,m_W,m_a)$ are the first derivatives of
$I(p^2,m_W,m_a)$ and $J(p^2,m_W,m_a)$ with respect to $p^2$, evaluated at
$p^2=m_i^2$, and 
the function $N(m_W,m_i,m_a,\xi_W)$ is defined in
Appendix~\ref{sec:appendix}.

The UV-divergent part of Eq.~(\ref{wfr:lepton}) is then
\begin{equation}
\Delta{\cal M}_{ij}^{{\rm leg,wfr,div}}=\frac{g^2}{32\pi^2}\overline{u}_i(p)
\sum_a B_{ia}B_{aj}^\dagger\left[\frac{m_im_j}{2m_W^2}a_-+
\left(\xi_W+\frac{m_a^2}{2m_W^2}\right)a_+\right]\Delta_W.
\end{equation}

If Eq.~(\ref{wfr:lepton}) is inserted in the leptonic $W$-boson decay
amplitude, important simplifications take place, in analogy with the analysis in Ref.~\cite{ref:1}. In fact, using Eqs.~(\ref{cond:B})--(\ref{cond:mass}) one readily finds that the contributions of the terms not involving $I'$ and $J'$ reduce to expressions that combine naturally with the proper vertex diagrams of Fig.~\ref{proper:diag}, an important property to ensure the cancellation of UV divergences and gauge dependences in the full physical amplitude.  
Although the corresponding contributions from the terms involving $I'$ and $J'$ do not simplify, we note that they are UV finite and gauge independent. 

\subsection{Outgoing Majorana neutrino}
\label{sec:majorana-SE}

\begin{figure}[t]
\begin{center}
\includegraphics{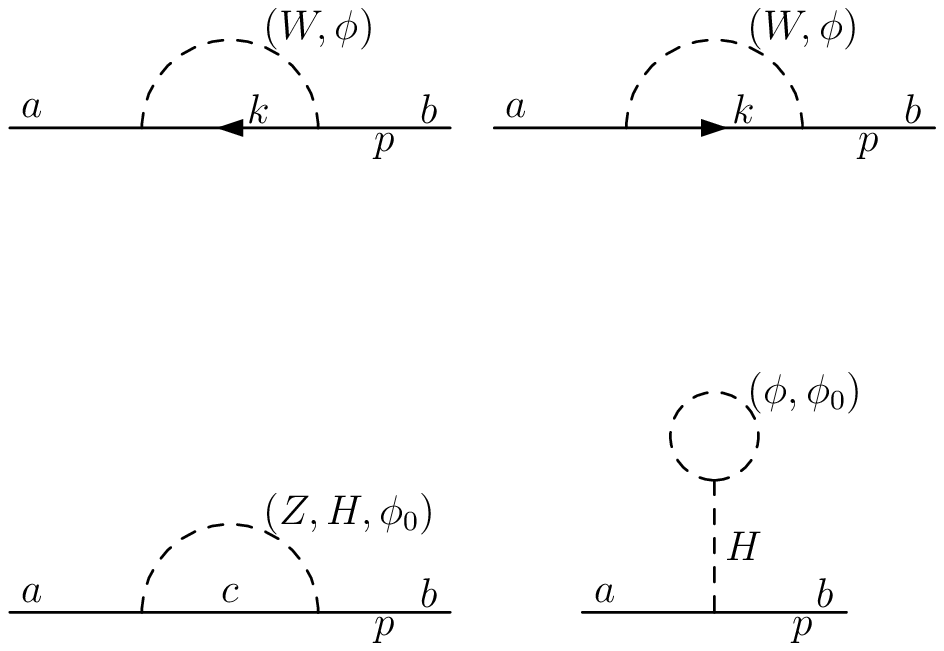}
\caption{Majorana neutrino self-energy diagrams.}
\label{majorana:diag}
\end{center}
\end{figure}

While the charged leptons could be treated analogously to the quarks,
the Majorana neutrinos require a more comprehensive analysis. 
In this case, mixing occurs not only in charged-current but also
in neutral-current interactions. 
For this reason, it is necessary to consider the corrections in Fig.~\ref{majorana:diag} induced by neutral currents, as well as those emerging from the charged currents, already present in the
charged-lepton case. 
Note that, if neutrinos were Dirac particles, the second diagram would be
absent.

As before, we evaluate the contributions in Fig.~\ref{majorana:diag} in the
$R_\xi$ gauges. 
The tadpole diagrams are needed to remove the gauge dependence in the diagonal
($aa$) and in parts of the non-diagonal ($ab$) contributions to the self-energy diagrams. 

We consider the case in which $a$ is an outgoing on-shell Majorana neutrino
and $b$ is a virtual Majorana neutrino. 
In analogy to the charged-lepton case, the sm contributions
$\Delta {\cal M}_{ab}^{\rm leg,sm}$ to the external-leg correction for an
outgoing on-shell Majorana neutrino $a$ read:
\begin{equation}
\Delta{\cal M}_{ab}^{{\rm leg,sm}}=\overline{u}_a(p)\left\{\displaystyle \frac{g^2}{32\pi^2}T_{ab}^{\rm sm}\right\} \frac{1}{\notp-m_b},
\label{sm:majorana}
\end{equation}
with
\allowdisplaybreaks{
\begin{eqnarray}
 T_{ab}^{\rm sm}&=& \left(m_aC^*_{ab}+m_bC_{ab}\right)
\left(a_++\frac{m_am_b}{2m_W^2}\Delta_Wa_-\right)\nonumber\\ 
&&{}+\left(m_aC_{ab}+m_bC^*_{ab}\right)\left(a_-+\frac{m_am_b}{2m_W^2}
\Delta_Wa_+\right)\nonumber\\ 
&&{}-\sum_k \frac{m_k^2}{2m_W^2}\left[
\left(m_aB_{ka}B^*_{kb}+m_bB^*_{ka}B_{kb}\right)a_+
+\left(m_aB^*_{ka}B_{kb}+m_bB_{ka}B^*_{kb}\right)a_-\right]\nonumber\\ 
&&{}\times\left[3\Delta_W+I(m_a^2,m_W,m_k)+J(m_a^2,m_W,m_k)\right]\nonumber\\
&&{}+\sum_k \left[\left(m_aB_{ka}B^*_{kb}+m_bB^*_{ka}B_{kb}\right)
\left(a_++\frac{m_am_b}{2m_W^2}a_-\right)\right.\nonumber\\ 
&&+\left.\left(m_aB^*_{ka}B_{kb}+m_bB_{ka}B^*_{kb}\right)
\left(a_-+\frac{m_am_b}{2m_W^2}a_+\right)\right]\nonumber\\ 
&&{}\times\left[I(m_a^2,m_W,m_k)-J(m_a^2,m_W,m_k)\right] \nonumber\\
&&{}+\frac{1}{4m_W^2}\sum_c\bigg[\left(m_aa_-+m_ba_+\right)
\left(m_am_bC_{ab}^*+5m_c^2C_{ac}C_{cb}\right)\nonumber\\
&&{}+\left(m_aa_++m_ba_-\right)\left(m_am_bC_{ab}+5m_c^2C_{ac}^*C_{cb}^*
\right)\nonumber\\
&&{}+4m_c^3\left(C_{ac}C_{cb}^*a_++C_{ac}^*C_{cb}a_-\right)\bigg]\Delta_H
\nonumber\\
&&{}+\frac{1}{4m_W^2}\sum_c \big[\left(m_aC^*_{ac}+m_cC_{ac}\right)
\left(m_cC_{cb}+m_bC^*_{cb}\right)\left(m_aa_-+m_ba_+\right)\nonumber\\
&&{}+\left(m_aC_{ac}+m_cC^*_{ac}\right)\left(m_cC^*_{cb}+m_bC_{cb}\right)
\left(m_aa_++m_ba_-\right)\big]\nonumber\\
&&{}\times\left[I(m_a^2,m_H,m_c)-J(m_a^2,m_H,m_c)\right]\nonumber\\
&&{}+\sum_c \frac{m_c}{2m_W^2}\big[\left(m_aC^*_{ac}+m_cC_{ac}\right)
\left(m_cC^*_{cb}+m_bC_{cb}\right)a_+\nonumber\\
&&{}+\left(m_aC_{ac}+m_cC^*_{ac}\right)\left(m_cC_{cb}+m_bC^*_{cb}\right)
a_-\big] I(m_a^2,m_H,m_c)\nonumber\\
&&{}+\frac{1}{2c_w^2}\left[C^*_{ab}+\frac{m_am_b}{2m_Z^2}C_{ab}\Delta_Z\right]
\left(m_aa_++m_ba_-\right) \nonumber\\
&&{}+\frac{1}{2c_w^2}\left[C_{ab}+\frac{m_am_b}{2m_Z^2}C^*_{ab}\Delta_Z\right]
\left(m_aa_-+m_ba_+\right) \nonumber\\
&&{}-\frac{m_c^3}{m_W^2}\Delta_Z\left(C_{ac}C_{cb}^*a_++C_{ac}^*C_{cb}a_-
\right) \nonumber\\
&&{}+\frac{1}{2c_w^2}\sum_c \left\{\left(1-\frac{m_c^2}{2m_Z^2}\right)
\left[C_{ac}C_{cb}(m_aa_-+m_ba_+)+C_{ac}^*C_{cb}^*(m_aa_++m_ba_-)\right]
\right.\nonumber\\
&&{}+\frac{1}{2m_Z^2}(m_aa_-+m_ba_+)(m_aC_ {ac}^*+m_cC_{ac})
(m_bC_{cb}^*+m_cC_{cb}) \nonumber\\
&&{}+\left.\frac{1}{2m_Z^2}(m_aa_++m_ba_-)(m_aC_ {ac}+m_cC_{ac}^*)
(m_bC_{cb}+m_cC_{cb}^*)\right\} \nonumber\\
&&{}\times \left[I(m_a^2,m_Z,m_c)-J(m_a^2,m_Z,m_c)\right] \nonumber\\
&&{}-\sum_c \frac{m_c^2}{4m_W^2}\left[C_{ac}C_{cb}\left(m_aa_-+m_ba_+\right)
+C_{ac}^*C_{cb}^*\left(m_aa_++m_ba_-\right)\right]\nonumber\\
&&{}\times \left[3\Delta_Z+I(m_a^2,m_Z,m_c)+J(m_a^2,m_Z,m_c)\right]\nonumber\\
&&{}+\sum_c \frac{m_c}{2c_w^2}\left[\left(4-\frac{m_c^2}{m_Z^2}\right)
\left(C_{ac}C_{cb}^*a_++C_{ac}^*C_{cb}a_-\right)\right.\nonumber\\
&&{}-\left.\frac{m_am_b}{m_Z^2}\left(C_{ac}^*C_{cb}a_++C_{ac}C_{cb}^*a_-
\right)\right]I(m_a^2,m_Z,m_c).
\label{sm:majorana2}
\end{eqnarray}}
The $I$ and $J$ functions, together with the UV divergences $\Delta_W$,
$\Delta_Z$, and $\Delta_H$ are defined in Appendix~\ref{sec:appendix}.

Equations~(\ref{sm:majorana})--(\ref{sm:majorana2}) are much lengthier than Eq.~(\ref{sm:lepton})
because of the additional class of diagrams considered. 
However, their structure is similar.

The wfr contributions $\Delta {\cal M}_{ab}^{\rm leg,wfr}$ to the external-leg
correction for an outgoing on-shell Majorana neutrino $a$ read:
\begin{equation}
\Delta {\cal M}_{ab}^{\rm leg,wfr}  = \overline{u}_a(p)
\left\{\frac{g^2}{32\pi^2}T_{ab}^{\rm wfr}\right\}, 
\label{wfr:majorana}
\end{equation}
with
\allowdisplaybreaks{
\begin{eqnarray}
T_{ab}^{\rm wfr}&=&\left(C_{ab}a_++C_{ab}^*a_-\right)
\left[1+\xi_W\left(\Delta_W-\frac{1}{2}+\frac{1}{2}\ln\xi_W\right)\right]
\nonumber\\
&&{}+\frac{1}{2c_w^2}(C_{ab}a_++C_{ab}^*a_-)\left[1+\xi_Z
\left(\Delta_Z-\frac{1}{2}+\frac{1}{2}\ln\xi_Z\right)\right]\nonumber\\
&&{}-\sum_k \left(B_{ka}^*B_{kb}a_++B_{ka}B_{kb}^*a_-\right)
N(m_W,m_a,m_k,\xi_W)\nonumber\\
&&-\frac{1}{2c_w^2}\sum_c (C_{ac}C_{cb}a_++C_{ac}^*C_{cb}^*a_-)
N(m_Z,m_a,m_c,\xi_Z)\nonumber\\
&&{}+\frac{1}{2c_w^2}\sum_c m_am_c
\left(C_{ac}C_{cb}^*a_-+C_{ac}^*C_{cb}a_+\right)M(m_Z,m_a,m_c,\xi_Z)\nonumber\\
&&{}+\frac{1}{2m_W^2}\sum_k \left[B_{ka}^*B_{kb}\left(m_am_ba_-+m_k^2a_+\right)
+B_{ka}B_{kb}^*\left(m_am_ba_++m_k^2a_-\right)\right]\nonumber\\
&&{}\times\left[\Delta_W+I(m_a^2,m_W,m_k)-J(m_a^2,m_W,m_k)\right]\nonumber\\
&&{}+\sum_k \left[B_{ka}^*B_{kb}a_++B_{ka}B_{kb}^*a_-\right]
\left[I(m_a^2,m_W,m_k)-J(m_a^2,m_W,m_k)\right]\nonumber\\
&&{}+\frac{1}{4m_W^2}\left(m_am_bC_{ab}^*+
\sum_c m_c^2C_{ac}C_{cb}\right)\Delta_Ha_+\nonumber\\
&&{}+\frac{1}{4m_W^2}\left(m_am_bC_{ab}+
\sum_c m_c^2C_{ac}^*C_{cb}^*\right)\Delta_Ha_-\nonumber\\
&&{}+\frac{1}{4m_W^2}\sum_c \left[(m_aC_{ac}^*+m_cC_{ac})
(m_cC_{cb}+m_bC_{cb}^*)a_+\right.\nonumber\\
&&{}+\left.(m_aC_{ac}+m_cC_{ac}^*)(m_cC_{cb}^*+m_bC_{cb})a_-\right]\nonumber\\
&&{}\times\left[I(m_a^2,m_H,m_c)-J(m_a^2,m_H,m_c)\right]\nonumber\\
&&{}+\frac{1}{2c_w^2}\sum_c (C_{ac}C_{cb}a_++C_{ac}^*C_{cb}^*a_-)
\left[I(m_a^2,m_Z,m_c)-J(m_a^2,m_Z,m_c)\right]\nonumber\\
&&{}+\frac{1}{4m_W^2}\sum_c \left[(m_aC_{ac}+m_cC_{ac}^*)
(m_cC_{cb}^*+m_bC_{cb})a_-\right.\nonumber\\
&&{}+\left.(m_aC_{ac}^*+m_cC_{ac})(m_cC_{cb}+m_bC_{cb}^*)a_+\right]\nonumber\\
&&{}\times\left[\Delta_Z+I(m_a^2,m_Z,m_c)-J(m_a^2,m_Z,m_c)\right]\nonumber\\
&&{}-\delta_{ab}\sum_k \frac{m_a^2m_k^2}{m_W^2}B_{ka}^*B_{kb}
\left[I'(m_a^2,m_W,m_k)+J'(m_a^2,m_W,m_k)\right]\nonumber\\
&&{}+2\delta_{ab}m_a^2\left(1+\frac{m_a^2}{2m_W^2}\right)\sum_k B_{ka}^*B_{kb}
\left[I'(m_a^2,m_W,m_k)-J'(m_a^2,m_W,m_k)\right]\nonumber\\
&&{}+\delta_{ab}\frac{m_a^2}{2m_W^2}\sum_c (m_aC_{ac}+m_cC_{ac}^*)
(m_cC_{cb}^*+m_bC_{cb})\nonumber\\
&&{}\times\left[I'(m_a^2,m_H,m_c)-J'(m_a^2,m_H,m_c)\right]\nonumber\\
&&{}+\delta_{ab}\sum_c \frac{m_am_c}{2m_W^2}\left[(m_aC_{ac}^*+m_cC_{ac})
(m_cC_{cb}^*+m_bC_{cb})a_+\right.\nonumber\\
&&{}+\left.(m_aC_{ac}+m_cC_{ac}^*)(m_cC_{cb}+m_bC_{cb}^*)a_-\right]
I'(m_a^2,m_H,m_c)\nonumber\\
&&{}+\delta_{ab}\frac{m_a^2}{c_w^2}\sum_c \left[\frac{1}{2m_Z^2}
(m_aC_{ac}+m_cC_{ac}^*)(m_bC_{cb}+m_cC_{cb}^*)\right.\nonumber\\
&&{}+\left.C_{ac}C_{cb}\left(1-\frac{m_c^2}{2m_Z^2}\right)\right]
\left[I'(m_a^2,m_Z,m_c)-J'(m_a^2,m_Z,m_c)\right]\nonumber\\
&&{}-\delta_{ab}\sum_c \frac{m_a^2m_c^2}{2m_W^2}C_{ac}C_{cb}
\left[I'(m_a^2,m_Z,m_c)+J'(m_a^2,m_Z,m_c)\right]\nonumber\\
&&{}+\delta_{ab}\sum_c \frac{m_am_c}{2c_w^2}
\left[\left(4-\frac{m_c^2}{m_Z^2}\right)
(C_{ac}C_{cb}^*a_++C_{ac}^*C_{cb}a_-)\right.\nonumber\\
&&{}-\left.\frac{m_a^2}{m_Z^2}(C_{ac}C_{cb}^*a_-+C_{ac}^*C_{cb}a_+)\right]
I'(m_a^2,m_Z,m_c).
\end{eqnarray}}

The UV-divergent part of Eq.~(\ref{wfr:majorana}) is:
\begin{eqnarray}
\Delta {\cal M}_{ab}^{\rm leg,wfr,div}&=& \frac{g^2}{32\pi^2}
\overline{u}_a(p)\bigg\{ (C_{ab}a_++C_{ab}^*a_-)
\left(\xi_W\Delta_W+\frac{1}{2c_w^2}\xi_Z\Delta_Z\right)\nonumber\\
&&{}+\frac{m_am_b}{2m_W^2}\left(C_{ab}^*a_++C_{ab}a_-\right)
\left(\Delta_W+\Delta_H\right)\nonumber\\
&&{}+\sum_k \frac{m_k^2}{2m_W^2}\left(B_{ka}^*B_{kb}a_++B_{ka}B_{kb}^*a_-
\right)\Delta_W \nonumber\\
&&{}+\sum_c \frac{m_c^2}{4m_W^2}\left(C_{ac}C_{cb}a_++C_{ac}^*C_{cb}^*a_-
\right)(\Delta_H+\Delta_Z)\bigg\}.
\end{eqnarray}

The discussion of the wfr contributions at the end of
Section~\ref{sec:charged-lepton-SE} remains valid. 
In fact, when inserted into the leptonic $W$-boson decay amplitude,
Eq.~(\ref{wfr:majorana}) is also subject to important simplifications. 
One then finds that the wfr contributions to the external leg involving
an outgoing Majorana neutrino can be combined naturally with the
proper vertex diagrams of Fig.~\ref{proper:diag}. 
To prove this, one needs to multiply Eq.~(\ref{wfr:majorana}) on the right by
$\left(-ig/\sqrt{2}\right)B_{ib}^*\gamma^\mu a_-v_i\varepsilon_\mu$, where
$v_i$ is the spinor associated with the charged lepton $l_i$ and
$\varepsilon_\mu$ is the polarization four-vector of the $W$ boson, and to
perform the summation over the index $b$. 
Making use of Eqs.~(\ref{cond:B})--(\ref{cond:mass}), it can
be verified that the terms in Eq.~(\ref{wfr:majorana}) not involving
derivatives of the amplitudes $I$ and $J$ lead to the structures
\begin{itemize}
\item[(i)] $B_{ia}^*f(m_i,m_a)$;
\item[(ii)] $\sum_b B_{ib}^*\left(C_{ba}f_1(m_i,m_a,m_b)
+C_{ba}^*f_2(m_i,m_a,m_b)\right)$.
\end{itemize}
The terms having the structure (i) combine naturally with
the proper vertex diagrams of Fig.~\ref{proper:diag}(c),
while those having the structure (ii) are to be combined with the diagrams depicted in Figs.~\ref{proper:diag}(a) and (b).
We emphasize that, also here, these terms include all the
gauge-dependent and UV-divergent contributions in Eq.~(\ref{wfr:majorana}). 
However, in Eq.~(\ref{wfr:majorana}) there are also terms proportional to
derivatives of the functions $I$ and $J$, which cannot be further simplified,
but are UV finite and gauge independent.

\subsection{Incoming leptons}
\label{incominglepton}

Equations (\ref{sm:lepton}), (\ref{wfr:lepton}), (\ref{sm:majorana}), and (\ref{wfr:majorana}) exhibit the sm and wfr contributions to the external-leg corrections in the case of an outgoing on-shell lepton. For the purpose of the following discussion, it is convenient to call $x$ and $y$ the flavors of the outgoing and virtual leptons. Thus, in Fig.~\ref{charged-lepton:diag}, $x=i$ and $y=j$, while in Fig.~\ref{majorana:diag}, $x=a$ and $y=b$.

The corresponding expressions for an incoming lepton of flavor $y$ is obtained by multiplying the diagrams in Figs.~\ref{charged-lepton:diag} and \ref{majorana:diag}, by $u_y(p)$ on the right and $(\notp-m_x)^{-1}$ on the left, where $x$ now denotes the virtual lepton. It is then easy to see that the sm contributions in the incoming case are obtained by interchanging $a_+\leftrightarrow a_-$ and $m_x\leftrightarrow m_y$ between the curly brackets of Eqs.~(\ref{sm:lepton}) and (\ref{sm:majorana}), and multiplying the resulting expression by $u_y(p)$ on the right and $(\notp-m_x)^{-1}$ on the left. Similarly, the wfr contributions for an incoming lepton of flavor $y$ are obtained by interchanging $a_+\leftrightarrow a_-$ and $m_x\leftrightarrow m_y$ between the curly brackets of Eqs.~(\ref{wfr:lepton}) and (\ref{wfr:majorana}), and multiplying the resulting expression by $u_y(p)$ on the right.

\section{Mass renormalization}
\label{sec:mass-count}

In this section, we study the cancellation of the sm contributions by suitably
adjusting the mass counterterms. 
We start with the simpler case of the charged leptons, which is, up to the
particle content, identical to that of quarks \cite{ref:1}. 
When treating the Majorana-neutrino case, a new feature appears.
One needs to keep in mind the fact that we are dealing with Majorana
particles, {\it i.e.}\ particles and antiparticles are identical. 
As a consequence, a new condition for the mass counterterms arises.

\subsection{Charged-lepton mass counterterm matrix}
\label{sec:leptoncountmass}

In order to generate mass counterterms suitable for the renormalization of the
sm contributions shown in Eq.~(\ref{sm:lepton}), we proceed as in
Ref.~\cite{ref:1}, where the case of quark mixing was considered. 
Decomposing the mass matrix as $m_0^{\prime l}=m^{\prime l}+\delta m^{\prime l}$,
where $m^{\prime l}$ and $\delta m^{\prime l}$ denote the renormalized and
mass counterterm matrices, and considering a bi-unitary transformation of the
charged-lepton fields $l^\prime_{L,R}$ that diagonalizes $m^{\prime l}$, the
mass term in the Lagrangian density takes the form
\begin{equation}
-\overline{l}(m^l+\delta m^{l(-)}a_-+\delta m^{l(+)}a_+)l,
\label{charged-lepton:mass-term}
\end{equation}
where $m^l$ is real, diagonal, and positive, and $\delta m^{l(-)}$ and
$\delta m^{l(+)}$ are arbitrary non-diagonal matrices subject to the
Hermiticity condition
\begin{equation}
\delta m^{l(+)}=\delta m^{l(-)\dagger}.
\label{hermiticity:lepton}
\end{equation}
The contribution of the mass counterterm to the external-leg amplitude is
given by
\begin{equation}
-i\overline{u}_i(\delta m_{ij}^{l(-)}a_-+\delta m_{ij}^{l(+)}a_+)
\frac{i}{\notp-m_j}.
\label{masscount:contribution}
\end{equation}

We now adjust $\delta m_{ij}^{l(-)}$ and $\delta m_{ij}^{l(+)}$ to cancel, as
much as possible, the sm contributions in Eq.~(\ref{sm:lepton}). 
The cancellation of the UV-divergent part is achieved by choosing
\begin{eqnarray}
\left(\delta m_{\rm div}^{l(-)}\right)_{ij}&=&-\frac{g^2m_i}{64\pi^2m_W^2}
\Delta_W\left(\delta_{ij}m_i^2-3\sum_a B_{ia}B_{aj}^\dagger m_a^2\right),
\nonumber\\
\left(\delta m_{\rm div}^{l(+)}\right)_{ij}&=&-\frac{g^2m_j}{64\pi^2m_W^2}
\Delta_W\left(\delta_{ij}m_i^2-3\sum_a B_{ia}B_{aj}^\dagger m_a^2\right).
\end{eqnarray}
Note that
\begin{equation}
 \left(\delta m_{\rm div}^{l(+)}\right)_{ij}
=\left(\delta m_{\rm div}^{l(-)}\right)_{ji}^*,
\end{equation}
so that the Hermiticity condition is fulfilled.

We call $ij$ channel the amplitude in which $i$ labels the outgoing on-shell
charged lepton and $j$ the virtual one. 
The $ji$ channel is then the amplitude in which the roles are reversed: $j$ is
the outgoing on-shell charged lepton, while $i$ is the virtual one. 
On the basis of Eq.~(\ref{masscount:contribution}), we define the mass
counterterms $\delta m_{ij}^{l(\pm)}$ such that they completely cancel the sm
corrections in Eq.~(\ref{sm:lepton}) for an outgoing charged lepton in the
$ij$ channel. 
As a consequence, we may write
\begin{eqnarray}
\delta m^{l(-)}_{ij}&=&-\frac{g^2m_i}{32\pi^2}\left\{\delta_{ij}
\left(1+\frac{m_i^2}{2m_W^2}\Delta_W\right)\right.\nonumber\\
&&{}+\sum_a B_{ia}B_{aj}^\dagger\left(1+\frac{m_j^2}{2m_W^2}\right)
\left[I(m_i^2,m_W,m_a)-J(m_i^2,m_W,m_a)\right]\nonumber\\
&&-\left.\sum_a B_{ia}B_{aj}^\dagger\frac{m_a^2}{2m_W^2}
\left[3\Delta_W+I(m_i^2,m_W,m_a)+J(m_i^2,m_W,m_a)\right]\right\},
\nonumber\\
\delta m^{l(+)}_{ij}&=&-\frac{g^2m_j}{32\pi^2}\left\{\delta_{ij}
\left(1+\frac{m_i^2}{2m_W^2}\Delta_W\right)\right.\nonumber\\
&&{}+\sum_a B_{ia}B_{aj}^\dagger\left(1+\frac{m_i^2}{2m_W^2}\right)
\left[I(m_i^2,m_W,m_a)-J(m_i^2,m_W,m_a)\right]\nonumber\\
&&{}-\left.\sum_a B_{ia}B_{aj}^\dagger\frac{m_a^2}{2m_W^2}
\left[3\Delta_W+I(m_i^2,m_W,m_a)+J(m_i^2,m_W,m_a)\right]\right\}.
\label{lepton-countmass}
\end{eqnarray}

Once $\delta m_{ij}^{l(-)}$ and $\delta m_{ij}^{l(+)}$ are fixed, the mass
counterterms for the reverse $ji$ channel are determined by the Hermiticity
conditions
\begin{equation}
\delta m_{ji}^{l(-)}=\delta m_{ij}^{l(+)*},\qquad
\delta m_{ji}^{l(+)}=\delta m_{ij}^{l(-)*}.
\end{equation}

We note that the functions $I$ and $J$ in Eq.~(\ref{lepton-countmass}) are
evaluated at $p^2=m_i^2$ in the $ij$ channel and at $p^2=m_j^2$ in the $ji$
channel. 
As a consequence, the mass counterterms cannot completely remove the sm
contributions in both amplitudes. 
Due to this restriction, we choose $\delta m_{ii}^l$ to cancel, as is
customary, all the diagonal contributions in Eq.~(\ref{sm:lepton}), while for
the non-diagonal entries, we choose $\delta m_{ij}^l$ with $i<j$ to cancel the
corresponding sm contributions. 
Once $\delta m_{ij}^l$ with $i<j$ are fixed, the mass counterterms for the $ji$ channel, {\em i.e.} $\delta m_{ji}^l$, are fixed
by the hermiticity conditions in Eq.~(\ref{hermiticity:lepton}).

\subsection{Majorana-neutrino mass counterterm matrix}
\label{sec:majoranacountmass}

In the weak-eigenstate basis, the bare mass matrix $m^{\prime 0,\nu}$ for the
neutrinos is symmetric and non-diagonal, and the corresponding terms in the
Lagrangian density are given in Eq.~(\ref{lagrangian:nu}). Decomposing $m^{\prime 0,\nu}=m^{\prime\nu}+\delta m^{\prime\nu}$, where
$m^{\prime\nu}$ and $\delta m^{\prime\nu}$ denote the renormalized and
counterterm mass matrices, we envisage a unitary transformation of the
Majorana-neutrino fields that diagonalizes $m^{\prime\nu}$, leading to a
renormalized mass matrix ${m}^{\nu}$ that is diagonal, real, and positive. 
As shown in Appendix~\ref{sec:appendix2}, this can be achieved by the
following transformation:
\begin{equation}
W^T{m'}^\nu W=m^\nu,
\end{equation} 
where $W$ is unitary. 
This also transforms $\delta m^{\prime \nu}$ into a new symmetric matrix
$\delta {m}^{\nu}$, which, in general, is non-diagonal. 
In the new framework, the mass term becomes
\begin{equation}
-\frac{1}{2}\overline{\nu}(m^\nu+\delta m^{\nu(-)}a_-+\delta m^{\nu(+)}a_+)\nu,
\end{equation}
where $m^\nu$ is real, diagonal, and positive, and $\delta m^{\nu(-)}$ and
$\delta m^{\nu(+)}$ are symmetric non-diagonal matrices subject to the
constraint
\begin{equation}
\delta m^{\nu(+)}=\delta m^{\nu(+)T}=\delta m^{\nu(-)*}
=\delta m^{\nu(-)\dagger}.
\label{hermiticity}
\end{equation}

As is customary, the mass counterterms are included in the interaction
Lagrangian density.
Their contribution to the external-leg amplitude reads:
\begin{equation}
 -i\overline{u}_a(p)(\delta m^{\nu(-)}_{ab}a_-+\delta m^{\nu(+)}_{ab}a_+)
\frac{i}{\notp-m_b}.
\label{count:majorana}
\end{equation}
We now adjust $\delta m^{\nu(-)}_{ab}$ and $\delta m^{\nu(+)}_{ab}$ to cancel,
as much as possible, the sm contributions given in Eq.~(\ref{sm:majorana}). 
The cancellation of the UV-divergent parts is achieved by choosing
\begin{eqnarray}
\left(\delta m_{\rm div}^{\nu(-)}\right)_{ab}&= & -\frac{g^2}{64\pi^2m_W^2}
\bigg\{m_am_b(m_aC_{ab}^*+m_bC_{ab})\left(\Delta_W+\frac{1}{2}\Delta_H
+\frac{1}{2}\Delta_Z\right)\nonumber\\
&&{}-3\sum_k m_k^2(m_aB_{ka}^*B_{kb}+m_bB_{ka}B_{kb}^*)\Delta_W\nonumber\\
&&{}+\sum_c m_c^2(m_aC_{ac}C_{cb}+m_bC_{ac}^*C_{cb}^*)
\left(\frac{5}{2}\Delta_H-\frac{3}{2}\Delta_Z\right)\nonumber\\
&&{}+2\sum_c m_c^3C_{ac}^*C_{cb}(\Delta_H-\Delta_Z)\bigg\},\nonumber\\
\left(\delta m_{\rm div}^{\nu(+)}\right)_{ab}&=&-\frac{g^2}{64\pi^2m_W^2}
\bigg\{m_am_b(m_aC_{ab}+m_bC_{ab}^*)\left(\Delta_W+\frac{1}{2}\Delta_H
+\frac{1}{2}\Delta_Z\right)\nonumber\\
&&{}-3\sum_k m_k^2(m_aB_{ka}B_{kb}^*+m_bB_{ka}^*B_{kb})\Delta_W\nonumber\\
&&{}+\sum_c m_c^2(m_aC_{ac}^*C_{cb}^*+m_bC_{ac}C_{cb})
\left(\frac{5}{2}\Delta_H-\frac{3}{2}\Delta_Z\right)\nonumber\\
&&{}+2\sum_c m_c^3C_{ac}C_{cb}^*(\Delta_H-\Delta_Z)\bigg\}.
\end{eqnarray}
It is easy to check that
\begin{equation}
\left(\delta m_{\rm div}^{\nu(+)}\right)_{ab}
=\left(\delta m_{\rm div}^{\nu(+)}\right)_{ba}
=\left(\delta m_{\rm div}^{\nu(-)}\right)_{ab}^*
=\left(\delta m_{\rm div}^{\nu(-)}\right)_{ba}^*,
\end{equation}
so that $\delta m^{\nu(-)}_{\rm div}$ and $\delta m^{\nu(+)}_{\rm div}$
satisfy the requirements in Eq.~(\ref{hermiticity}).

In order to discuss the cancellation of the UV-finite parts, as we did in the
charged-lepton case, we call $ab$ channel the amplitude in which $a$ labels
the outgoing on-shell Majorana neutrino and $b$ the virtual one. 
In the $ab$ channel, we define then the mass counterterms
$\delta m_{ab}^{\nu(\pm)}$ such that they fully cancel the sm contributions of
Eq.~(\ref{sm:majorana}) and obtain:
\begin{eqnarray}
-\frac{32\pi^2}{g^2}\delta m_{ab}^{\nu(-)}&=& -\frac{32\pi^2}{g^2}
\left(\delta m_{\rm div}^{\nu(-)}\right)_{ab}+\left(m_aC_{ab}+m_bC^*_{ab}
\right)\left(1+\frac{1}{2c_w^2}\right)\nonumber\\ 
&&{}-\sum_k \frac{m_k^2}{2m_W^2}\left(m_aB^*_{ka}B_{kb}+m_bB_{ka}B^*_{kb}
\right)\nonumber\\ 
&&{}\times\left[I(m_a^2,m_W,m_k)+J(m_a^2,m_W,m_k)\right]\nonumber\\
&&{}+\sum_k \left[\frac{m_am_b}{2m_W^2}\left(m_aB_{ka}B^*_{kb}
+m_bB^*_{ka}B_{kb}\right)
+\left(m_aB^*_{ka}B_{kb}+m_bB_{ka}B^*_{kb}\right)\right]\nonumber\\ 
&&{}\times\left[I(m_a^2,m_W,m_k)-J(m_a^2,m_W,m_k)\right] \nonumber\\ 
&&{}+\frac{1}{4m_W^2}\sum_c \big[m_a\left(m_aC^*_{ac}+m_cC_{ac}\right)
\left(m_cC_{cb}+m_bC^*_{cb}\right)\nonumber\\
&&{}+m_b\left(m_aC_{ac}+m_cC^*_{ac}\right)\left(m_cC^*_{cb}+m_bC_{cb}\right)
\big]\nonumber\\
&&{}\times\left[I(m_a^2,m_H,m_c)-J(m_a^2,m_H,m_c)\right]\nonumber\\
&&{}+\sum_c \frac{m_c}{2m_W^2}\left(m_aC_{ac}+m_cC^*_{ac}\right)
\left(m_cC_{cb}+m_bC^*_{cb}\right)I(m_a^2,m_H,m_c)\nonumber\\
&&{}+\frac{1}{2c_w^2}\sum_c \left\{\left(1-\frac{m_c^2}{2m_Z^2}\right)
\left(m_aC_{ac}C_{cb}+m_bC_{ac}^*C_{cb}^*\right)\right.\nonumber\\
&&{}+\frac{m_a}{2m_Z^2}(m_aC_ {ac}^*+m_cC_{ac})(m_bC_{cb}^*+m_cC_{cb})
\nonumber\\
&&{}+\left.\frac{m_b}{2m_Z^2}(m_aC_ {ac}+m_cC_{ac}^*)(m_bC_{cb}+m_cC_{cb}^*)
\right\} \nonumber\\
&&{}\times \left[I(m_a^2,m_Z,m_c)-J(m_a^2,m_Z,m_c)\right] \nonumber\\
&&{}-\sum_c \frac{m_c^2}{4m_W^2}\left(m_aC_{ac}C_{cb}
+m_bC_{ac}^*C_{cb}^*\right)\nonumber\\
&&{}\times \left[I(m_a^2,m_Z,m_c)+J(m_a^2,m_Z,m_c)\right] \nonumber\\
&&{}+\sum_c \frac{m_c}{2c_w^2}\left[\left(4-\frac{m_c^2}{m_Z^2}\right)
C_{ac}^*C_{cb}-\frac{m_am_b}{m_Z^2}C_{ac}C_{cb}^*\right] I(m_a^2,m_Z,m_c),
\nonumber\\
-\frac{32\pi^2}{g^2}\delta m_{ab}^{\nu(+)}&=&-\frac{32\pi^2}{g^2}
\left(\delta m_{\rm div}^{\nu(+)}\right)_{ab}+\left(m_aC^*_{ab}+m_bC_{ab}
\right)\left(1+\frac{1}{2c_w^2}\right)\nonumber\\ 
&&{}-\sum_k \frac{m_k^2}{2m_W^2}\left(m_aB_{ka}B^*_{kb}+m_bB^*_{ka}B_{kb}
\right)\nonumber\\
&&{}\times\left[I(m_a^2,m_W,m_k)+J(m_a^2,m_W,m_k)\right]\nonumber\\
&&{}+\sum_k \left[\frac{m_am_b}{2m_W^2}\left(m_aB^*_{ka}B_{kb}
+m_bB_{ka}B^*_{kb}\right)+\left(m_aB_{ka}B^*_{kb}+m_bB^*_{ka}B_{kb}
\right)\right]\nonumber\\ 
&&{}\times\left[I(m_a^2,m_W,m_k)-J(m_a^2,m_W,m_k)\right] \nonumber\\ 
&&{}+\frac{1}{4m_W^2}\sum_c \big[m_b\left(m_aC^*_{ac}+m_cC_{ac}\right)
\left(m_cC_{cb}+m_bC^*_{cb}\right)\nonumber\\
&&{}+m_a\left(m_aC_{ac}+m_cC^*_{ac}\right)\left(m_cC^*_{cb}+m_bC_{cb}\right)
\big]\nonumber\\
&&{}\times\left[I(m_a^2,m_H,m_c)-J(m_a^2,m_H,m_c)\right]\nonumber\\
&&{}+\sum_c \frac{m_c}{2m_W^2}\left(m_aC^*_{ac}+m_cC_{ac}\right)
\left(m_cC^*_{cb}+m_bC_{cb}\right)I(m_a^2,m_H,m_c)\nonumber\\
&&{}+\frac{1}{2c_w^2}\sum_c \left\{\left(1-\frac{m_c^2}{2m_Z^2}\right)
\left(m_bC_{ac}C_{cb}+m_aC_{ac}^*C_{cb}^*\right)\right.\nonumber\\
&&{}+\frac{m_b}{2m_Z^2}(m_aC_ {ac}^*+m_cC_{ac})(m_bC_{cb}^*+m_cC_{cb})
\nonumber\\
&&{}+\left.\frac{m_a}{2m_Z^2}(m_aC_ {ac}+m_cC_{ac}^*)(m_bC_{cb}+m_cC_{cb}^*)
\right\} \nonumber\\
&&{}\times \left[I(m_a^2,m_Z,m_c)-J(m_a^2,m_Z,m_c)\right] \nonumber\\
&&{}-\sum_c \frac{m_c^2}{4m_W^2}\left(m_bC_{ac}C_{cb}+m_aC_{ac}^*C_{cb}^*
\right)\nonumber\\
&&{}\times \left[I(m_a^2,m_Z,m_c)+J(m_a^2,m_Z,m_c)\right] \nonumber\\
&&{}+\sum_c \frac{m_c}{2c_w^2}\left[\left(4-\frac{m_c^2}{m_Z^2}\right)
C_{ac}C^*_{cb}-\frac{m_am_b}{m_Z^2}C^*_{ac}C_{cb}\right] I(m_a^2,m_Z,m_c).
\label{majorana-countmass}
\end{eqnarray}

Also here, the functions $I$ and $J$ are evaluated at $p^2=m_a^2$ in the $ab$
channel and at $p^2=m_b^2$ in the $ba$ channel. 
Therefore, the mass counterterms in Eq.~(\ref{majorana-countmass}) cannot completely
remove the sm contributions of Eq.~(\ref{sm:majorana}) in both channels. 
We then choose $\delta m_{aa}^\nu$ to cancel all the diagonal
contributions in Eq.~(\ref{sm:majorana}) and $\delta m_{ab}^\nu$ with $a<b$ to
fully cancel the corresponding sm contributions. 
Once $\delta m_{ab}^{\nu}$ with $a<b$ are fixed, the mass counterterms for the
$ba$ channel, {\em i.e.} $\delta m_{ba}^{\nu}$, are determined by the
conditions in Eq.~(\ref{hermiticity}).

\section{Renormalization of mixing matrices}
\label{sec:mixing-count}

In the previous section, we have shown how one can define mass counterterms on
the basis of the sm contributions calculated in Section~\ref{sec:self-energies}. 
In particular, in both charged-lepton and Majorana-neutrino cases, the
UV-divergent parts in the sm contributions of Eqs.~(\ref{sm:lepton}) and
(\ref{sm:majorana}) are completely canceled by the mass counterterms. 
In addition, also UV-finite parts get canceled, up to the Hermiticity conditions
(\ref{hermiticity:lepton}) and (\ref{hermiticity}). 
We wish to emphasize that the mass counterterms constructed in this way are
explicitly gauge independent.

In what follows, we proceed with the diagonalization of the complete mass
matrices, which include the renormalized and counterterm mass matrices. 
Similar to the quark case \cite{ref:1}, this procedure leads to mixing
matrix counterterms which automatically satisfy the basic properties
(\ref{cond:B})--(\ref{cond:mass}) and are gauge independent. 
As before, we first discuss the case of charged leptons followed by that of
Majorana neutrinos.

The renormalized fermion masses thus resulting are the familiar on-shell
masses, which coincide with the pole masses \cite{Kniehl:2008cj} to the order
of our calculation.

\subsection{Diagonalization of the charged-lepton mass matrix}

Following Ref.~\cite{ref:1}, we consider a bi-unitary transformation that diagonalizes the complete charged-lepton mass matrix in Eq.~(\ref{charged-lepton:mass-term}) through terms of ${\cal O}(g^2)$. Calling $U^l_L$ and $U^l_R$ the unitary matrices in this transformation and writing
\begin{equation}
U^l_L=1+ih^l_L,\qquad U^l_R=1+ih^l_R,
\label{eq:Ul}
\end{equation}
where $h^l_L$ and $h^l_R$ are hermitian matrices of ${\cal O}(g^2)$, one finds that the off-diagonal elements ($i\not = j$) are given by
\begin{eqnarray}
i(h^l_L)_{ij}&=&-\frac{m^l_i\delta m^{l(-)}_{ij}+\delta m^{l(+)}_{ij}m^l_j}
{(m^l_i)^2-(m^l_j)^2},
\nonumber\\
 i(h^l_R)_{ij}&=&-\frac{m^l_i\delta m^{l(+)}_{ij}+\delta m^{l(-)}_{ij}m^l_j}
{(m^l_i)^2-(m^l_j)^2},
\label{lepton:h}
\end{eqnarray}
while the diagonal elements can be chosen to vanish, namely $(h_{L,R}^l)_{ii}=0$. As shown in Appendix B in Ref.~\cite{ref:1}, the alternative choice $(h_{L,R}^l)_{ii}\not = 0$ has no physical effect on the $Wl\nu$ interactions through ${\cal O}(g^2)$. 

\subsection{Diagonalization of the Majorana-neutrino mass matrix}

The situation in the case of Majorana neutrinos is similar to the one of the
charged leptons, except that now one needs only one unitary matrix for the
diagonalization of the complete mass matrix.
Writing
\begin{equation}
U^\nu=1+ih^\nu,
\label{eq:Um}
\end{equation}
where $h^\nu$ is again a Hermitian matrix of ${\cal O}(g^2)$, one finds that the off-diagonal elements ($a\not =b$) are given by
\begin{equation}
i(h^\nu)_{ab}=-\frac{m^\nu_a\delta m^{\nu(-)}_{ab}
+\delta m^{\nu(+)}_{ab}m^\nu_b}{(m^\nu_a)^2-(m^\nu_b)^2},
\label{majorana:h}
\end{equation}
and, in analogy with the charged-lepton case, we choose $(h^\nu)_{aa}=0$.

\subsection{Mixing counterterm matrices}

We analyze next the effect of the transformations of Eqs.~(\ref{eq:Ul})--(\ref{majorana:h})
on the $Wl\nu$ coupling in Eq.~(\ref{Wlnu:int}). 
Performing the above transformations, we find through terms of
${\cal O}(g^2)$ that
\begin{equation}
{\cal L}_W=-\frac{g}{\sqrt{2}}W_\mu^-\overline{l}(B+\delta B)\gamma^\mu
a_-\nu +{\rm h.c.},
\end{equation}
where
\begin{equation}
\delta B=i(Bh^\nu-h^l_LB).
\end{equation}
It is easy to verify that both the renormalized and bare mixing matrices
satisfy the first condition in Eq.~(\ref{cond:B}) while, due to the second
condition, once $\delta B$ is fixed, $\delta C$ is fixed as well, leading to
\begin{equation}
\delta C=i(Ch^\nu-h^\nu C).
\end{equation}
One can further check that all the other conditions in Eqs.~(\ref{cond:C}) and
(\ref{cond:mass}) for the two mixing matrices are satisfied. 
Of course, all the equalities hold through the order of the calculation, namely ${\cal O}(g^2)$.

For completeness, we give the two counterterm matrices in component form:
\begin{eqnarray}
\displaystyle\delta B_{ia} & =&i\left[\sum_b B_{ib}
\left(h^\nu\right)_{ba}-\sum_j \left(h^l_L\right)_{ij}B_{ja}\right]\nonumber\\
& =&-\sum_b B_{ib}\frac{m^\nu_b\delta m^{\nu(-)}_{ba}
+\delta m^{\nu(+)}_{ba}m^\nu_a}{(m^\nu_b)^2-(m^\nu_a)^2}
+\sum_j \frac{m^l_i\delta m^{l(-)}_{ij}+\delta m^{l(+)}_{ij}m^l_j}
{(m^l_i)^2-(m^l_j)^2}B_{ja},
\label{B:count}\\
\delta C_{ab} &=&i\sum_c \left[C_{ac}\left(h^\nu\right)_{cb}
-\left(h^\nu\right)_{ac}C_{cb}\right]\nonumber\\
&=&-\sum_c C_{ac}\frac{m^\nu_c\delta m^{\nu(-)}_{cb}
+\delta m^{\nu(+)}_{cb}m^\nu_b}{(m^\nu_c)^2-(m^\nu_b)^2}
+\sum_c \frac{m^\nu_a\delta m^{\nu(-)}_{ac}+\delta m^{\nu(+)}_{ac}m^\nu_c}
{(m^\nu_a)^2-(m^\nu_c)^2}C_{cb},
\label{C:count}
\end{eqnarray}
where $\delta m_{ij}^{l(\pm)}$ and $\delta m_{ab}^{\nu(\pm)}$ are the off-diagonal
mass counterterms determined in Sections~\ref{sec:leptoncountmass} and
\ref{sec:majoranacountmass}, respectively, 
and it is understood that $b\not =a$ in the first and $j\not =i$ in the second
term of Eq.~(\ref{B:count}) and that $c\not =b$ in the first and $c\not =a$ in
the second term of Eq.~(\ref{C:count}).

Since the mass counterterms are adjusted to cancel the off-diagonal sm
contributions to the extent allowed by the properties of the mass matrices,
the same is true of the mixing counterterm matrices $\delta B$ and $\delta C$. 
In particular, they fully cancel the UV-divergent parts of the off-diagonal sm
contributions.

\section{Conclusions}
\label{sec:conclusions}

In this paper, we have generalized the on-shell framework to renormalize the CKM
matrix at the one-loop level proposed in Ref.~\cite{ref:1} to extensions of the SM that include Majorana neutrinos, an appealing scenario that may explain the smallness of the observed neutrino masses and may lead to neutrino-less double beta decays. The presence of Majorana neutrinos requires a separate analysis, due to
modified interactions and symmetry factors leading to a generically different
set of Feynman rules. 
Here, the mixing generally also occurs in neutral-current interactions. 
However, once the Feynman rules are established, the procedure is similar to
the case of the CKM matrix.

We showed how gauge-independent mass counterterms can be fixed by means of the
sm contributions and how they lead to mixing counterterm matrices. 
We gave explicit expressions for $\delta B$ and $\delta C$. 
They are consistent with the properties satisfied by the two mixing matrices
and are explicitly gauge independent. 
We saw that once $\delta B$ is fixed, $\delta C$ is fixed as well, as a
consequence of the second property in Eq.~(\ref{cond:B}). 
However, one could also choose to fix the $\delta C$ counterterm separately,
{\it e.g.}\ by choosing to study the $Z\nu\nu$ coupling, with the same result.

\section*{Acknowledegments}

B.A.K. and A.S. are grateful to the Max Planck Institute for Physics in Munich
for the warm hospitality during a stay when part of this work was carried out.
This work was supported in part by the German Research Foundation through the
Collaborative Research Center No.~676 {\it Particles, Strings and the Early
Universe --- the Structure of Matter and Space Time}.  
The work of A. Sirlin was supported in part by the National Science Foundation
through Grant Nos.\ PHY--0245068 and PHY--0758032.

\appendix
\section{Definitions}
\label{sec:appendix}

In this appendix, we gather important definitions used throughout this work.
The UV divergences which appear in the expressions of the sm and wfr
contributions and later in the mass and mixing counterterm matrices are
defined by
\begin{equation}
\Delta_B=\frac{1}{n-4}+\frac{1}{2}\left(\gamma_E-\ln 4\pi\right)
+\ln\frac{m_B}{\mu}\equiv\Delta+\ln\frac{m_B}{\mu}, 
\label{defdelta}
\end{equation}
where $n$ is the space-time dimension, $\gamma_E$ the Euler's constant, $\mu$
is the 't~Hooft mass scale, and $m_B$ is the mass of boson $B=W,Z,H$.

The functions $I$, $J$, $N$, and $M$ are defined through the integrals:
\begin{eqnarray}
\{I;J\}(p^2,m_1,m_2)&=&\int_0^1{\rm d}x\{1;x\}
\ln\frac{m_2^2x+m_1^2(1-x)-p^2x(1-x)-i\varepsilon}{m_1^2},
\nonumber\\
N(m_1,m_2,m_3,\xi_1)&=&\frac{1}{m_1^2}\int_0^1{\rm d}x
\left[m_2^2(1-x)+m_3^2\right]\nonumber\\
&&{}\times\ln\frac{m_3^2x+m_1^2\xi_1(1-x)-m_2^2x(1-x)-i\varepsilon}
{m_3^2x+m_1^2(1-x)-m_2^2x(1-x)-i\varepsilon},\nonumber\\
M(m_1,m_2,m_3,\xi_1)&=&\frac{1}{m_1^2}\int_0^1{\rm d}x\,x
\ln\frac{m_3^2x+m_1^2\xi_1(1-x)-m_2^2x(1-x)-i\varepsilon}
{m_3^2x+m_1^2(1-x)-m_2^2x(1-x)-i\varepsilon}.
\end{eqnarray}

The above integrals are not all independent.
In fact, the integrals $J$, $N$, and $M$ can be expressed by means of $I$ as
\begin{eqnarray}
J(p^2,m_1,m_2)&=&\frac{1}{2p^2}\left[-m_2^2+m_1^2+m_2^2
\ln\frac{m_2^2}{m_1^2}+(p^2-m_2^2+m_1^2)\,I(p^2,m_1,m_2)\right],
\nonumber\\
M(m_1,m_2,m_3,\xi_1)&=&\frac{1}{m_1^2}\left[J(m_2^2,m_1\sqrt\xi_1,m_3)
-J(m_2^2,m_1,m_3)+\ln\xi_1\right],
\nonumber\\
N(m_1,m_2,m_3,\xi_1)&=&\frac{m_2^2+m_3^2}{m_1^2}
\left[I(m_2^2,m_1\sqrt\xi_1,m_3)-I(m_2^2,m_1,m_3)+\ln\xi_1\right]\nonumber\\
&&{}-m_2^2\,M(m_1,m_2,m_3,\xi_1).
\label{Ndef}
\end{eqnarray}
Note that these integrals represent UV-finite parts of the standard scalar
one-loop integrals \cite{'tHooft:1978xw}.
In fact, we have
\begin{equation}
I(p^2,m_1,m_2)=-2\Delta_1-B_0(p^2,m_1,m_2),
\end{equation}
where $\Delta_1$ is defined by Eq.~(\ref{defdelta}) and $B_0$ is defined as in
Ref.~\cite{ref:19}.
According to Eq.~(\ref{Ndef}), the integrals $J$, $N$, and $M$ can be written
in terms of the scalar two-point function $B_0$ as well.

\section{Majorana-neutrino mass matrix diagonalization}
\label{sec:appendix2}

According to the singular-value decomposition theorem, any complex matrix
$M^\prime$ can be diagonalized by a bi-unitary transformation of the form
\begin{equation}
M=S^\dagger M^\prime U,
\label{diagproof1}
\end{equation}
where $S$ and $U$ are unitary matrices and $M$ is real and diagonal with
non-negative eigenvalues.
In the proof of Eq.~(\ref{diagproof1}), $S$ is chosen such that
\begin{equation}
S^\dagger M^\prime M^{\prime\dagger} S = M^2.
\label{diagproof2}
\end{equation}
For a symmetric matrix $M^\prime$, this becomes:
\begin{equation}
S^\dagger M^\prime M^{\prime *} S = M^2.
\label{diagproof3}
\end{equation}
We now take the hermitian adjoint of Eq.~(\ref{diagproof1}),
\begin{equation}
U^\dagger M^{\prime\dagger} S = M,
\label{diagproof4}
\end{equation}
multiply Eq.~(\ref{diagproof1}) on the left by Eq.~(\ref{diagproof4}), and
take the complex conjugate,
\begin{equation}
U^T M^\prime M^{\prime *} U^* = M^2.
\label{diagproof5}
\end{equation}
Comparing Eqs.~(\ref{diagproof3}) and (\ref{diagproof5}), we see that we can
identify
\begin{equation}
S=U^*,\qquad S^\dagger=U^T.
\label{diagproof6}
\end{equation}
Inserting Eq.~(\ref{diagproof6}) in Eq.~(\ref{diagproof1}) we obtain
\begin{equation}
M=U^T M^\prime U,
\label{diagproof7}
\end{equation}
which tells us that, for the diagonalization of a complex symmetric matrix, one
needs only one unitary transformation \cite{ref:20}.


\end{document}